\documentclass[mn2e,]{article}
\input{epsf}
\newcommand{\be}{\begin{eqnarray}}
\newcommand{\ee}{\end{eqnarray}}

\begin{document}

\title{Formation of the first halos, galaxies and
magnetic fields}
\maketitle

\author
\begin{center}
M. Demia\'nski$^{1,2}$,  A. Doroshkevich$^{3,4}$,
T. Larchenkova$^{3}$
\end{center}

\vspace{0.5cm}
1. Institute of Theoretical Physics, University
        of Warsaw,   02-093 Warsaw, Poland\\
2. Department of Astronomy, Williams College,
           Williamstown, MA 01267, USA\\	
3. Lebedev Physical Institute of  Russian Academy
  of Sciences, 119333 Moscow,  Russian Federation\\
4. National Research Center Kurchatov Institute,
  123182, Moscow, Russian Federation\\
\vspace{0.5cm}		

\section*{Abstract}
Cosmic objects with magnetic fields (quasars,
radiogalaxies) are observed at redshifts $z\geq 7$
(Wang et al., 2021, Fan et al., 2023, Yang et al.,
2024) and more (for instance,
at $z = 10.073\pm 0.002$, Goulding et al., 2023)
indicates the early creation of magnetic fields.
Observations of the cosmic telescope JWST show
that the first galaxies were formed at redshifts
$z\simeq$ 15 -- 20. We link the transformation of
the low mass dark matter halos into galaxies with the
creation of the first stars owing to the radiative
cooling of partly ionized hydrogen at $z\geq 10$.
The early formation of galaxies creates favourable
conditions for high impact of the Compton
scattering of the relic radiation photons
and electrons on the electron temperature
and leads to the partial separation of electrons and
protons. Together with turbulent motions such
separation stimulates creation of magnetic fields
on a galactic scale. The same processes distort the
primordial perturbations of the relic radiation, what
can be observed and confirmed in special simulations.

\section*{Introduction}

Despite the significant progress achieved in recent
years in the development of the standard cosmological
model (SCM) (McQuinn 2017, Naab \& Ostriker, 2017,
Tumlinson et al., 2017, Wechsler \& Tinker, 2018,
Salucci 2019, Zavala \& Frenk 2019, de Martino 2020)
the problem of emergence of the first stars and
magnetic fields in early Universe remains relevant
right up to now. A magnetic field can arise
on an appropriate scale during periods of inflation
and/or phase transitions (Subramanian 2019, Steinmetz
2023, Regis 2023) but for the formation
of the observed large-scale magnetic fields
the period of formation of the first stars and the
transformation of dark matter (DM) halos into observable
galaxies is most favourable. Recent observations of the
James Webb Telescope ~(JWST) (Menci 2022, Atek 2023,
Labbe 2023, Xiao 2023, Donnan 2023, McLeod 2023,
Baggenm 2023, Fujimoto 2023, Ormerod 2023, Finkelstein
2023, Castellano 2023, Wang 2024, Atek 2024)
and the ALMA radio observatory~(Fujimoto 2023)
show that these processes occur at $10\leq z\leq 20$
and lead to the reionization of the Universe.

Let us recall that according to the SCM, the development
of density and velocity perturbations leads to the
formation of DM halos and clouds of baryons with masses
$M_{halo}\sim 10^5 - 10^7~M_\odot$. Such halos are
predominantly concentrated in vast regions of high
density with masses $M_{halo}\sim 10^9 - 10^{12}~M_\odot$.
Subsequent hierarchical condensation and coagulation of
DM halos in these areas leads to the formation of
massive galaxies and observable elements of
structure of the Universe.

In  (Shull et al. 2012) 
it was noticed that the fraction of baryons in the
intergalactic medium $f_{IGM}$ is $\sim 0.8$.
Observations of Fast Radio Burst (Fortunato et al.,
2023, Lemos et al., 2023) confirm these estimates
\[
f_{IGM}\simeq 0.8(1\pm 0.1)\,.
\]
This means that galaxies and associated DM halos
contain only about 10 - 20\% of the mass of DM
and baryons, collected in DM halos with masses
$M_{halo}/M_\odot\sim 10^{5}-10^{12}$. In numerical
models (Wang 2023) fraction of baryons increase to
\[
f_{IGM}\simeq 0.93(1\pm 0.08)\,,
\]
and, accordingly, the fraction of baryons
included in the galaxies decreases.

These data show that the influence of galaxies on
the dynamics of the Universe is small. However,
observations and investigation of the properties and
evolution of galaxies allows us to get closer
to understanding the complex processes of formation
of the observable Universe, such as the emergence of
stars, magnetic fields, the ultraviolet background
and secondary ionization of the intergalactic medium.
Along with optical, infrared and radio observations,
analysis of the above processes require observations
in other ranges of the electromagnetic spectrum SKA
(Heald 2020), ALMA (Fujimoto et al., 2023), 
Millimetron (Novikov et al., 2021) 
etc. over the entire range of available redshifts.

\section{Parameters of low-mass galaxies}

From the point of view of understanding how the
large-scale magnetic fields were formed, most interesting
is the period of formation of the first stars and
transformation of low-mass DM halos into galaxies.
This implies the need to know the parameters
of these objects. Modern observations of early
(low-mass) galaxies are quite limited, so
in our analysis we will use observations of 8 low-
mass galaxies (and DM halos) of the Local Group
(Walker et al., 2009, Weisz et al., 2014, Golini et al.
2024). The main parameters of these objects are
presented in Table \ref{tbl8}.

\begin{table}
\caption{Parameters of galaxies and low-mass DM halos
  used is our analysis, data from (Walker et al., 2009)
}
\label{tbl8}
\vspace{4mm}
\begin {tabular}{llr rcc rccc c}
Name&$r_{1/2}$&$\sigma_v$&$M_{1/2}$&$\rho_{1/2}$
&$R_{DM}$&$M_{DM}$&$\rho_{DM}$\cr
&kpc&km/s&$10^6M_\odot$&$10^8M_\odot/kpc^3$&
$kpc$&$10^6M_\odot$&$10^8M_\odot/kpc^3$\cr
 Carina &0.14 & 6.6& 3.4& 3.2& 0.8& 8.4&0.04\cr
 Draco &0.22 & 9.1&11.0& 2.3& 1.8&57.6&0.03\cr
 Fornax &0.34 &11.7&27.0& 1.6& 1.7&18.5&0.01\cr
 Leo I &0.13 & 9.2& 6.5& 6.6& 1.0&16.2&0.05\cr
 Leo II &0.12 & 6.6& 3.1& 4.0& 0.4&13.1&0.02\cr
 Sculptor &0.09 & 9.2& 4.6&13.0& 1.1& 5.4&0.11\cr
 Sextans &0.29 & 7.9&11.0& 1.0& 1.0& 5.6&0.09\cr
 Ursa minor&0.15 & 9.5& 7.8& 5.5& 0.7&16.7&0.06\cr
\hline
 average &0.18 & 8.7& 9.3& 4.6& 1.1&17.7&0.04\cr

\end{tabular}

\end{table}

As follows from Fig. 1 in (Walker et al., 2009)
in these DM halos the velocity dispersion weakly
depends on radius, which fits well with the popular
NFW halo model (Navarro et al., 1997)
with the density profile
\be
\rho(r)=\frac{\rho_0}{x(1+x)^2},\quad x=\frac{r}{r_0},
\quad v^2=v_0^2\left[\frac{\ln(1+x)}{x}-\frac{1}{1+x}
\right]\,,
\label{rhoNFW}
\ee
with the sizes of galaxies $r_{1/2}$ and DM halos
$R_{DM}$ differ by a factor of $\langle R_{DM}/r_{1/2}
\rangle\sim 6-7$. This means that the masses of
galaxies and halos  also differ by a factor of
$\sim 6-7$, and their average densities differ by
a factor of $\sim 50 - 100$. These differences need
to be taken into account when discussing the
transformation of DM halos into galaxies.

A wide variety of early galaxies and DM halos is 
also supported by the existence of a dark galaxy
(Montes et al., 2024) with mass $M_{1/2}\simeq 4\cdot
10^8M_\odot$, radius $R_{1/2}\simeq 6.9$ kpc
and velocity dispersion $\sigma_v\simeq 30$ km/s.

Close values of the velocity dispersion in the DM
halos and in galaxies (see Table \ref{tbl8}) indicate
temperatures of the order of
\be
T_{gas}\simeq 10^4K\,.
\label{tgas}
\ee
Typical time of evolution of DM halos $t_{dm}$ and
galaxies $t_{gal}$, as well as the average density of
matter, can be estimated as
\be
t_{dm}=(4\pi G \rho_{dm})^{-1/2}\simeq 10^{15}s,\quad
\langle\rho_m\rangle=10^7M_\odot/kpc^3\,,
\label{cs}
\ee
\[
t_{gal}=(4\pi G \rho_{gal})^{-1/2}\simeq 2\cdot 10^{14}s,
\quad \langle\rho_m\rangle=4\cdot 10^8M_\odot/kpc^3\,.
\]
Both times $t_{dm}$ and $t_{gal}$ are small compared
to the cosmological time
\be
t_{cs}(z)\simeq 2\cdot 10^{16}z_{10}^{-3/2}s,\quad
\rho_m(z)=37(1+z)^3M_\odot/kpc ^3,\quad
z_{10}=(1+z)/10\,,
\label{cosmo}
\ee
therefore, galaxies and dark matter halos can be
considered as stationary objects.

Such DM halos could form at redshifts $z\sim 20 - 10$,
which is confirmed by JWST observations of
galaxies with masses $M_{gal}\sim 10^{10}M_\odot$ at
redshifts $z\sim 13 - 8$. Mergers of 
these DM halos into galaxies with masses $M_{gal}\geq
10^{10}M_\odot$ observed at low redshifts, occur at
$z\leq 6$ (Demianski et al., 2023).

\section{Transformation of DM halos into galaxies and
 formation of the first stars}

According to the SCM, transformation of DM halos into
low-mass galaxies and the formation of the first stars
occurs at redshifts $z\sim 15-10$. This transformation
is caused in part by the radiative cooling of ionized
baryons collected in the DM halos, it leads to their
condensation at the center of halos with maintaining
their temperature at $T_{gas}\sim 10^4K$ and increasing
density of the gas. Cooling is caused by bremsstrahlung
and recombination radiation, radiation in hydrogen and
helium lines etc. These processes are observed in
numerical models (Abel 2007). 
Due to the development of thermal instability
(isobaric mode) at this time the first stars are
rapidly formed and a stationary low-mass galaxy forms
in the central region of the DM halo (Schauer et al,
2021, Nakauchi et al., 2021, Sadanari 2021, Chon 2022).
The stability of such a galaxy is provided by the
joint pressure of the DM and recently formed stars
(Walker et al., 2009). Various scenarios for the
formation of the first stars are discussed, for
example, in (Klessen \& Glover 2023, Schauer et al.,
2023).  

In this work, star formation is assumed to be due to
the development of thermal instability in the process
of cooling and concentration of baryons in the center
of already (partially) formed DM halo that occurs at
almost constant temperature of DM particles and baryons
$T_{DM}\sim T_{gas}\sim 10^4K$. This allows us to consider
cooling of the gas as due to the conventional radiation
processes.

In  DM halos with the density profile 
(Navarro et al., 1997) and parameters $r_0$, $v_0$
the density of the (small) baryon fraction $\rho_b(r)$
corresponds to the isothermal distribution (Demianski
et al., 2022) with temperature $T_{gas}$ and speed of
sound $a_T$
\be
\rho_b(r)\simeq \rho_b(0)\exp[-\kappa_T^2(1-\ln(1+x)/x)],
\quad x=r/r_0\,,
\label{isoT}
\ee
\[
\kappa_T^2=v_0^2/a^2_T,\quad a_T^2=kT_{gas}/m_p\,,
\]
where $m_p$ is the proton mass. This profile is
maintained  at $\rho_b\leq \rho_{DM}$ and later is
transformed into the standard profile (\ref{rhoNFW})
(Navarro et al., 1997).

A detailed study of these complex processes requires
use of special numerical models. But
some important features of such a transformation,
leading to the formation of magnetic fields may be
estimated without detailed calculations. Note that
JWST observations of galaxies shift the process of
formation of dark matter halos and galaxies to
redshifts $z\sim 15 - 20$, on which an important role
plays Compton scattering of the cosmic microwave
background radiation on free electrons of partially
ionized plasma. This process sets the temperature of
the electrons equal to the temperature of the cosmic
microwave background radiation
\be
T_e\simeq 2.7(1+z)K,\quad z\geq 10\,,
\label{Te}
\ee
which is 200 times smaller than the temperature of
hydrogen and protons $T_{gas}\simeq 10^4 K$ in the
galaxies and DM halos.

According to the SCM, DM halos have a noticeable
angular moment (Peebles, 1969, Doroshkevich, 1970,
White, 1984, Demia\'nski et al., 2011, Kriel et al.
2022). In these conditions different temperature of
electrons and protons leads to their relative
motions what creates currents and magnetic fields
on a scale of stars and DM halos.

\subsection{Numerical estimates}

The cooling of electrons at $z\geq 10$ occurs
by scattering of photons of the cosmic microwave
background radiation (Zeldovich \& Novikov, 1983,
Shandarin \& Zeldovich, 1989) with a characteristic
time
\be
t_e=\frac{3m_ec}{8\sigma_T\varepsilon_\gamma}
\simeq 4\cdot 10^{15}z_{10}^{-4}s,\quad z_{10}=(1+z)/10\,,
\label{compton}
\ee
where $\sigma_T$ is the Compton scattering cross
section and $\varepsilon_{\gamma}\propto (1+z)^4$ is
the energy density of the cosmic microwave background
radiation. Comparison of times $t_{dm}$, $t_{gal}$ and
$t_{e}$ shows that when $z\geq 10$ this process
maintains the electrons temperature at
\be
T_{CMB}\simeq 2.7(1+z)K\,.
\label{tcmb}
\ee

The influence of the cosmic microwave background
radiation strongly depends on parameters of the DM
halo, it is more effective at redshifts $z_{eff}\geq
14$, and quickly decreases with time. This indicates
diversity of magnetic properties of galaxies that
form at different redshifts. Comparison of (\ref{cs})
and  (\ref{compton}) shows that for
$z\leq 10$ this influence is negligible and the
galaxies formed later do not have noticeable magnetic
fields.

Estimates of circular velocities ${\bf j}$ and angular
moment ${\bf J}$ of the DM halos at $z\leq 3$ were
obtained in numerical models (Demia\'nski et al., 2011):
\[
{\bf J}\sim f_J [{\bf R\sigma_v}],\quad
{\bf j}\sim f_j[{\bf R\sigma_v}],
\]
\be
\langle f_J\rangle\simeq 0.2,\quad
\langle f_j\rangle\simeq 0.8\,,
\label{moment}
\ee
where $R$ and $\sigma_v$ characterize the size and
velocity dispersion of halo. The distribution
functions of the random amplitudes  $f_J$ and $f_j$
are close to the exponential and Gaussian (Demia\'nski
et al., 2011). Using observations (see Table
\ref{tbl8}), one can estimate the dynamic parameters
of low-mass DM halos:
\[
\langle J\rangle\sim 2~kpc~km/s,\quad
\langle j\rangle\sim 8~kpc~ km/s\,,
\]
and for the magnetic moment
\be
{\bf \mu}\simeq \frac{e}{2mc}|{\bf J}|\,.
\label{magm}
\ee

These estimates confirm the effectiveness of the
proposed process. However, quantitative estimates of
emerging magnetic fields require more detailed modeling.

\section{Discussion}

At high redshifts $z\geq 7$
(Wang et al, 2021, Fan et al., 2023, Yang et al., 2024)
magnetic fields are observed in radio galaxies and
quasars (see, for example, $z = 10.073\pm 0.0027$
(Goulding et al., 2023)). These observations
indicate that on the scale of stars and galaxies,
magnetic fields already existed during the period of
reionization, but how they
were created is still unclear. Fields are discussed
arising during inflation, during phase transitions in
early Universe, with the annihilation of matter -
antimatter, with decay or annihilation of dark matter
particles, in cosmic shocks waves, with the possible
separation of protons and electrons by the radiation
of stars (Subramanian, 2019, Regis et al., 2023,
Steinmetz et al., 2023). On the
contrary, evolution of the magnetic fields is well
modeled (see, for example, Rodrigues et al., 2015,
Martin--Alvarez et al., 2021, Kriel et al., 2022, etc).

According to the SCM, the development of density and
velocity perturbations leads to formation of halos
with masses $M_{DM}\sim 10^6 - 10^7~M_\odot$, consisting
of 84\% of DM and only 16\% of baryons (Komatsu et al.,
2011, Ade et al., 2016). Evolution of the first
halos leads to ignition of the first stars,
transformation of the DM halos into galaxies and
reionization of the Universe. Recent JWST
observations  shift the formation redshifts
of DM halos and low-mass galaxies up to z = 15 -- 20,
which allows us to consider in our analysis interaction
of free electrons with the  cosmic microwave background
radiation. This interaction results in separation of
electrons and protons by temperature and density
(\ref{tgas}), (\ref{tcmb}).
Under these conditions, developed turbulence arising
in low-mass non-spherical halos DM (Peebles, 1969,
Doroshkevich, 1970, White, 1984, Demia\'nski et al.,
2011, Kriel et al. 2022) naturally leads to the
appearance of currents and magnetic fields on the
scale of stars, galaxies and DM halos (see, however,
Girichidis 2021). 

Interaction of the hot gas with the microwave background
radiation (Sunyaev Zeldovich effect, Sunyaev, Zeldovich
1972, Bahk \& Hwang 2024) alter properties
of the cosmic microwave background radiation. These
questions deserve further study both in observations,
and in special numerical models (Libeskind et al., 2020).
The transformation of a DM halo into a galaxy
is caused by the flow of cooled partially ionized
hydrogen towards the center of the halo accompanied with a
moderate restructuring of the DM density profile.
The deceleration of electrons leads to accumulation
of the positive charges in the central part of the halo
while maintaining the overall electrical neutrality of
the halo. This effect inhibits the flow of protons to the
center and concentration of mass in the center may occur
due to inflow to the center of neutral hydrogen and helium.
The over all density profile of the halo is determined by
superposition of distributions of DM, stars, gas,
protons and electrons.

This restructuring is accompanied by the formation of
the first stars due to thermal instability of the
cooling baryons (isobaric mode). Further transformation
 of the DM halo into the galaxy is accompanied by
increasingly stronger influence of stars. Observations
show (Walker et al., 2009) that the structure of
galaxies is determined by the motion of dark matter
and stars. It is important that all these processes
occur simultaneously, and the formation of periphery
of galaxies and DM halos is accompanied by explosions
of the first stars in the central region
(Walker et al., 2009, de Martino et al., 2020).
At redshifts $z\geq 4$ this promotes the formation
of $Ly_\alpha$ emitters along with LSB galaxies
(Novikov et al., 2021).

Obviously, quantitative estimates of the  transformation
of DM halos into galaxies depend on many factors and
require complex numerical simulations. Full reproduction
of these processes are possible only in special
numerical models (see Rodrigues 2015, Libeskind 2020,
Matin--Alvarez 2021, Nakauchi et al., 2021, Sadanari
et al., 2021, Chon et al, 2021). However, it is
important to note that recent JWST observations that
detected early massive galaxies, qualitatively changed
the usual picture of transformation of DM halos into
galaxies and allows us to incorporate the emergence
of the large-scale magnetic fields into the general
analysis of the process of formation of DM halos and
galaxies.


This work was carried out within the framework of the
FIAN NIS program 36-2024.

{}


\begin{thebibliography}{}

\bibitem[1]{abel20}
  Abel, T, ApJ, 659, L87 (2007)

\bibitem[2]{atek23}
Atek, H., Chemerinska, I., Wang, B., et al., 2023,  24
MNRAS, 524, 54

\bibitem[3]{atek24}
Atek, H., Labbe, I., Furtak, L., et al., 2024,
arXiv:2308.08540

\bibitem[4]{ayr22}
Ayromlou, M., Nelson, D., Pillepich, A.,
2022, MNRAS, arXiv:2211.07659

\bibitem[5]{bagg23}
Baggenm F., van Dokkum, P., Labbe, I., et al., 2023,
ApJ, 955, L12

\bibitem[6]{bh24}
Bahk, H., Hwang, H., 2024,  ApJS, 272, 7 

\bibitem[7]{castel23}
Castellano, M., Fontana, A., Treu, T., et al.,
2023,  ApJ., 948, 14   

\bibitem[8]{raffa22}
Chon, S., Ono, H., Omukai, K., Schneider, R.), 2022,
MNRAS, 514, 4639 

\bibitem[9]{martino20}
de Martino, I., Chakrebarty, S., Cesare, V.,
Gallo, A., Ostorero, L., Diaferio, A., 2020,
Universe, 6, 107

\bibitem[10]{ddps11}
Demia\'nski, M., Doroshkevich, A., Pilipenko, S.,
Gottlober, S. 2011, MNRAS, 414, 1813

\bibitem[11]{ddl22}
Demianski, M., Doroshkevich, A., Larchenkova, T., 2022,
Astron.Letters, 48, 361

\bibitem[12]{ddl23}
Demia\'nski, M., Doroshkevich, A., Larchenkova, T.,
2023, Sov.A.  67, 439  

\bibitem[13]{D23}
Donnan, C., McLeod, D., McLure, R., et al., 2023,
MNRAS, 520, 4554

\bibitem[14]{d70}
Doroshkevich, A., 1970, Astrophysics, 6, 320

\bibitem[15]{fan23}
Fan, X., Banados, E., Simcoe, R., 2023, Ann. Rev.,
  61, 373

\bibitem[16]{fink23}
Finkelstein, S., Bagley, M., Ferguson, H., et al.,
2023, ApJ, 946, L13    

\bibitem[17]{fort23}
Fortunato, J., Hipolito-Ricardi, W., dos Santos,  M.,
2023, arXiv:2307.04711  

\bibitem[18]{f23}
Fujimoto, S., Bezanson, R., Labbe, I., et al., 2023,
arXive: 2309.07834 

\bibitem[19]{giri21}
Girichidis. P., 2021, MNrAS, 507, 5641

\bibitem[20]{gal24}
Golini, G, Mancera Pina P., Trujillo, I., Montes, M., 2024,
arXiv:2404.06537 
  
\bibitem[21]{goulding23}
Goulding, A., Greene, J., Setton, D., et al.,
2023, ApJL, 955, L24

\bibitem[22]{ska20}
Heald, G., Mao, S., Vacca, V., et al., 2020, Galaxies,
8, 53

\bibitem[23]{klessen23}
Klessen, R., \& Glover, S., Ann. Rev. 2023, \textbf{61}, 65 

\bibitem[24]{koma11}
Komatsu, E., et al., 2011, ApJS, 182, 18

\bibitem[25]{kriel22}
Kriel, N., Beattie, J., Seta, A., Federrath, C., 2022,
MNRAS, 513, 2457  

\bibitem[26]{l23}
Labbe, I., van Dokkum, P., Nelson, E., et al. 2023,
Nature, 616, 266L  

\bibitem[27]{lem23}
Lemos, T., Goncalves, R., Carvallo, J., Alcaniz, J.,
2023, arXiv:2307.06911

\bibitem[28]{lib20}
Libeskind, N., Carlesi, E., Grand, R., et al., 2020,
MNRAS, 498, 2968

\bibitem[29]{katz21}
Matin--Alvarez, S., Katz, H., Sijacki, D., Devriendt, J.,
Slyz, A., 2021, MNRAS, 504, 2517

\bibitem[30]{M23}
McLeod, D., Donnan, C., McLure, R., et al., 2023,
arXiv:2304.14469

\bibitem[31]{mcq16}
McQuinn, M., 2016, ARA\&A, 54, 313

\bibitem[32]{M22}
Menci, N., 2022, ApJ, 938,

\bibitem[33]{montes24}
Montes, M., Trujillo, I., et al. 2024, A\&A, 681, A15 

\bibitem[34]{naab17}
Naab, T., \& J.Ostriker, J., 2017 Ann. Rev. \textbf{55}, 59

\bibitem[35]{omu21}
Nakauchi, D., Omukai, K., Suza, H., 2021,  MNRAS, 502, 3394

\bibitem[36]{nfw}
Navarro, J., Frenk, C., White, S.,  1997, MNRAS, 275,
720; ApJ, 490, 493

\bibitem[37]{ufn21}
Novikov, I., Likhachev, S., Shchekinov, Yu., et al.,
2021, Physics -- Uspekhi, 64, 386

\bibitem[38]{ormer23}
Ormerod, K., Conselice, C., Adams, N., et al., 2023,
MNRAS, 527, 611

\bibitem[39]{ttt}
Peebles, P.J.E., 1969, APJ, 155, 393

\bibitem[40]{regis23}
Regis, M., Korsmeier, M., Bernardi, G., et al., 2023, 
JCAP, 8, 30  

\bibitem[41]{shukurov15}
Rodrigues, L., Shukurov, A., Fletcher, A., Baugh, C.,
2015, MNRAS, 450, 3472

\bibitem[42]{sada21}
Sadanari, K., Omukai, K., Sugimura, K., Matsumoto,
T., Tomida, K., 2021,  MNRAS, 505, 4197
  
\bibitem[43]{salu19}
Salucci, P., 2019, A\&ARv \textbf{27}, 2

\bibitem[44]{Schauer23}
Schauer, A., Boylan-Kolchin, M., Colston K., Sameie O.,
Bromm V., Bullock J.S., Wetzel A., 2023, ApJ, 950, 20 

\bibitem[45]{sz89}
Shandarin, S. \& Zeldovich, Ya., 1989, RvMP, 61, 185

\bibitem[46]{shull12}
Shull, J., Smith, B., Danforth, C., 2012, ApJ, 759, 23

\bibitem[47]{stein23}
Steinmetz, A., Yang, C., Rafelski, J., 2023, PhRvD,
108I3522 

\bibitem[48]{sub19}
Subramanian, K., 2019, Galaxies, 7, 47; 

\bibitem[49]{sz72}
Sunyaev, R. \& Zeldovich, Ya., 1972, Comments on
Astrophysics and Space Physics, 4, 173

\bibitem[50]{tuml17}
Tumlinson, J., Peebles, M., Werk, J., 2017, Ann. Rev.
\textbf{55}, 389

\bibitem[51]{walker09}
Walker, M., Mateo, M., Olszewski, E., et al. 2009,
ApJ., 704, 1274

\bibitem[52]{wang21}
Wang, F., Yang,J., Fan, X., et al., 2021, ApJL, 907, L1

\bibitem[53]{wang231}
Wang, B., Wei, J., 2023, ApJ, 944, 50

\bibitem[54]{wang230}
Wang, B., Leja, J., Labbe, I., at all.,2023,
arXiv:2310.0127

\bibitem[55]{wechs}
Wechsler, R., \& Tinker, J., 2018, Ann.Rev.\textbf{56},
435

\bibitem[56]{weisz14}
Weisz, D., et al., 2014, ApJ, 789, 24;

\bibitem[57]{wtt}
White, S.D.M., 1984, ApJ, 286, 38

\bibitem[58]{X23}
Xiao, M., Oesch, P., Elbaz, D., et al., 2023,
arXiv:2309.02492

\bibitem[59]{Yang24}
Yang, D., Schindler, J., Nann, R., et al., 2024, MNRAS,
528, 2679 

\bibitem[60]{zasov17}
Zasov, A. B., Saburova, A. C.,  Khoperskov, A. B.,
Khoperskov, C. A.,  2017, Phys.Usp. \textbf{60}, 3

\bibitem[61]{zawala19}
Zavala, T., \& Frenk, C., 2019, Galaxy \textbf{7},
81 

\bibitem[62]{zn83}
Zeldovich, Ya., Novikov, D., 1983,
Relativistic Astrophysis, Vol. 2 Structure and Evolution
of the Universe, The Chicago University  Press, Chicago.

\end{thebibliography}
\end{document}